\documentclass[aps,preprint]{revtex4}
\begin{document}
\title{ On the anomalous extreme ultraviolet emission lines in helium-hydrogen 
plasma }
\author{S.C. Tiwari}
\affiliation{Institute of Natural Philosophy \\1, Kusum Kutir, Mahamanapuri,
Varanasi - 221 005, India } 
\begin{abstract}
The idea of photonic de Broglie waves is discussed in the context of recent
obsrved lines in the helium-hydrogen plasma.
\end{abstract}
\maketitle
The reported observations \cite{1} of new emission lines in the 
helium-hydrogen (98-2 \%) plasmas have been given a simple explanation 
in \cite{2} . The reported observations show sharp emission lines with 
the energies satisfying empirical relations given by
\begin{equation}
E_e = 13.6 n \; {\rm eV} \qquad \qquad  {\rm where \; n = 1, 2, 3, 7, 9}
\end{equation}
\begin{equation}
E_e = 13.6 n - 21.21  \; {\rm eV} \qquad   {\rm where \; n = 4, 6, 8}
\end{equation}
Let us note that the ground state energy level of hydrogen is -13.6 eV, 
and this immediately hints at intriguing physics if the observations 
are correct. In view of the doubtful nature of 'exotic' observations in 
recent years  in the so-called cutting-edge research,we must be 
skeptical of this report unless independent experiments by other groups 
verify the observations. Assuming the correctness of the reported 
emission lines, let us see if Sathyamurthy \cite{2} offers any 
explanation. He says that, 'if there are several H atom recombinations 
taking place simultaneously, under favourable circumstances, there 
could be emissions corresponding to integral multiples of 13.6 eV..' 
Obviously this does not amount to any physical interpretation, and one 
does not know what it means to say 'several atom recombinations'.
   
Is there some possible mechanism to explain the observations if these are
true?  Let us explore the hypothesis of photonic de Broglie waves \cite{3} 
in this context. Jacobson et al in \cite{3} propose that an ensemble 
of photons in quantum optics could be considered as a Bose condensate 
with de Broglie wavelength $\lambda/N$ where $\lambda$ is the
wavelength, and $N$ the average number of the constituent photons. To give a 
simple derivation,let us consider N photons with frequency $\nu$ ,then the
total energy $Nh\nu$ can be used to obtain the momentum of the condensate
\begin{equation}
p= Nh\nu/c
\end{equation}
The de Broglie wavelength is 
\begin{equation}
\lambda_b= \frac{h}{p} = \frac{\lambda}{N}
\end{equation}
The idea that N photons with frequency $\nu$ could behave as a single entity
with frequency $N\nu$ seems strange,and in the literature counterintuitiveness 
of quantum mechanics is invoked.Experiments on biphoton interference and
quantum entanglement have been of active areas of research besides the
geometrical phases for such photon pairs \cite{4}.

Let us assume that helium is instrumental in creating cavities where photonic
condensates could form for the photons with energy 13.6 eV. One could then
explain relation(1). Following our work on geometrical phase \cite{4} we argue
that spin plays an important role in the condensate i.e. even numbers give
spinless state and odd number of photons give spin one state.The difference
between relations (1) and (2) is that in the former we have odd n (excepting 2)
while in (2) it is even. Here Sathyamurthy's suggestion that photon condensate
gets inelastically scattered from helium seems useful. He confines only to
energy conservation,however the angular momentum and parity selection rules
have to be properly incorporated in a fuller discussion.

If the condensate mechanism is correct one can look for its verification using
the Mach-Zehnder interferometer as is proposed by Ryff and Ribeiro in
\cite{4}.Here the source beam should be the emission radiation from the
plasma,and the interference pattern has to be contrasted with the independent
source at this frequency.

Obviously the idea of photonic de Broglie waves is speculative,and in spite of
intense work in quantum optics the problems at fundamental level lack
understanding. A radically new approach envisages photon as a composite
structure and a number of photons could coalesce to create photo-balls that
behave as photons \cite{5}. This approach also desreves attention.

{\it Acknowledgements}

I am grateful to Prof. Y. Singh, Physics Department,BHU Varanasi for drawing my
attention to Ref. 2, and for a lively discussion on this problem.

\end{document}